\documentclass[12pt]{article}
\usepackage{graphicx,color,amssymb,amsmath}

\usepackage[a4paper, total={16cm,22cm}]{geometry}

\begin{document}

\newcommand{\TODO}[1]{ {\color{blue} ( #1 )} }

\newcommand{\lrf}[2]{ \left(\frac{#1}{#2}\right)}
\newcommand{\lrfp}[3]{ \left(\frac{#1}{#2} \right)^{#3}}
\newcommand{\vev}[1]{\left\langle #1\right\rangle}

\newcommand{\TeV}{\text{TeV}}
\newcommand{\GeV}{\text{GeV}}
\newcommand{\MeV}{\text{MeV}}
\newcommand{\keV}{\text{keV}}
\newcommand{\eV}{\text{eV}}

\newcommand{\pbar}{\bar{p}}

\begin{titlepage}
\begin{flushright}
UT-15-14\\
IPMU-15-0055
\end{flushright}
\vskip 3cm
\begin{center}
{\Large \bf
AMS-02 Antiprotons from \\ Annihilating or Decaying Dark Matter 
}
\vskip 1.5cm
{
Koichi Hamaguchi, Takeo Moroi, Kazunori Nakayama
}
\vskip 0.9cm
{\it Department of Physics, University of Tokyo, Bunkyo-ku, Tokyo 113--0033, Japan \vspace{0.2cm}
\par
Kavli Institute for the Physics and Mathematics of the Universe (Kavli IPMU), \\
University of Tokyo, Kashiwa 277--8583, Japan
}
\vskip 2cm
\abstract{

Recently the AMS-02 experiment reported an excess of cosmic ray antiprotons 
over the expected astrophysical background.
We interpret the excess as a signal from annihilating or decaying dark matter 
and find that the observed spectrum is well fitted by adding 
contributions from the annihilation or decay of dark matter with mass of ${\cal O}$(TeV) or larger.
Interestingly, Wino dark matter with mass of around 3\,TeV, whose thermal relic abundance is consistent with present dark matter abundance,
can explain the antiproton excess.
We also discuss the implications for the decaying gravitino dark matter with R-parity violation.

}
\end{center}
\end{titlepage}

\section{Introduction}

The existence of the dark matter (DM) has been confirmed by various cosmological observations~\cite{Agashe:2014kda}, yet its identity is a complete mystery.
The DM provides the most robust evidence for physics beyond the Standard Model.

Recently, the AMS-02 collaboration has reported 
their latest results of the cosmic-ray antiproton measurement~\cite{AMS02}, which may be indirect signatures of annihilating/decaying DM in our Universe.
Although recent studies~\cite{Giesen:2015ufa,Evoli:2015vaa} have claimed that the AMS-02 antiproton flux is within the uncertainties of the astrophysical secondary antiproton flux, the predicted secondary antiproton flux still tends to be smaller than the observed one at higher energy $\gtrsim 100~\GeV$, which may indicate a DM contribution in that energy range. Refs.~\cite{Giesen:2015ufa,Evoli:2015vaa,Jin:2015sqa} also derived upper bound on the DM signal, but the conservative bound~\cite{Evoli:2015vaa} is still weak, which allows a large DM contribution. 

In this letter, we consider annihilating and decaying DM
as a possible source of the AMS-02 antiproton flux.
We show that annihilating/decaying DM with a mass of ${\cal O}(\TeV)$ can explain the antiproton flux in the high energy range.
We also investigate the implications for supersymmetric (SUSY) DM. 
It is shown that the AMS-02 antiprotons may originate from 
Wino DM with a mass of 2--3\,TeV.
Surprisingly, the Wino mass of around 3\,TeV, which is 
suitable for the thermal relic DM scenario, 
can explain the observed antiproton data.
Another interesting DM candidate is gravitino with an R-parity violation.
It is shown that ${\cal O}(\TeV)$ gravitino DM with an R-parity violation 
can also be the source of the AMS-02 antiprotons.

\section{Antiproton from annihilating and decaying DM}

The flux of primary antiprotons from DM annihilation/decay 
at the Solar System, $\vec{r}=\vec{r}_{\odot}$, 
is given by\footnote{We neglect the difference between the fluxes at the top of the Earth's atmosphere $\Phi_{\pbar}^{\text{TOA}}$ and at the interstellar $\Phi_{\pbar}^{\text{IS}}$ due to the solar modulation, since its effect is ${\cal O}(1)\%$ for $T\gtrsim 50~\GeV$.}
\begin{align}
\Phi_{\pbar}^{\text{DM}}(T) = \frac{v(T)}{4\pi} f_{\pbar}(T,\vec{r}_{\odot})\,,
\end{align}
where $v(T)$ is the velocity of the antiproton with kinetic energy $T$ and $f_{\pbar}(T,\vec{r})$ is the antiproton number density per unit kinetic energy. The propagation of antiprotons is described by a cylindrical stationary diffusion model~\cite{Maurin:2002ua}
\begin{align}
0
&=\frac{\partial}{\partial t}f_{\pbar}(T,\vec{r})
\nonumber
\\
&=
\nabla\left[ K(T)\nabla f_{\pbar}(T,\vec{r})\right] 
-
\frac{\partial}{\partial z}
\left[ \text{sign} (z) V_\text{c} f_{\pbar}(T,\vec{r}) \right]
-2h \delta(z) \Gamma_{\text{ann}}(T)f_{\pbar}(T,\vec{r}) 
+
Q(T,\vec{r}),
\label{eq:diff}
\end{align}
with boundary conditions $f_{\pbar}(T,\vec{r})=0$ at $r=R$ and $z=\pm L$, 
where $(r,\varphi,z)$ are the galactic cylindrical coordinates.
Here, the effects of energy losses, reacceleration, and tertiary antiprotons are neglected.

In Eq.~(\ref{eq:diff}), $K(T)$ is the diffusion coefficient and assumed to be spatially constant. 
It is parametrized as $K(T,\vec{r})=K(T)=K_0 \beta (p/\text{GeV})^\delta$, where
$p$ and $\beta=v(T)/c$ are the momentum and velocity of the antiproton, respectively.
The $V_\text{c}$ term represents the convective wind, which is assumed to be
constant and perpendicular to the galactic plane.
The third term represents the annihilation of the antiproton on 
interstellar protons in the galactic plane, where
$h$ represents the thickness of the galactic plane and 
$\Gamma_{\text{ann}}(T)=(n_\text{H}+4^{2/3}n_{\text{He}})\sigma^\text{ann}_{p\pbar} v(T)$ is the annihilation rate. 
We take $h=0.1$ kpc, $n_\text{H}=1~\text{cm}^{-3}$,  $n_{\text{He}}=0.07n_\text{H}$,
and $\sigma^\text{ann}_{p\pbar}$ given in Refs.~\cite{Tan:1984ha,Hisano:2005ec},
\begin{align}
\sigma^\text{ann}_{p\pbar} =
\begin{cases}
661(1+0.0115T^{-0.774}-0.948T^{0.0151})\text{mb} & T<15.5\text{GeV}\,,
\\
36 T^{-0.5}\text{mb} & T\ge 15.5\text{GeV}\,.
\end{cases}
\end{align}
Lastly, 
$Q(T,\vec{r})$ is the source term of the antiprotons. 
We adopt the set of propagation parameters $R$,
$L$,  $K_0$,  $\delta$,   and $V_\text{c}$ in Ref.~\cite{Donato:2003xg}, which are shown in Table~\ref{tab:paras}.

\begin{table}[t]
\begin{center}
\begin{tabular}{|c|ccccc|}
\hline
Model & $R$(kpc) & $L$(kpc) & $K_0$(kpc$^2$/Myr) & $\delta$ & $V_\text{c}$(km/s)
\\
\hline
MIN & 20 & 1 & 0.0016 & 0.85 & 13.5
\\
MED & 20 &4  & 0.0112 & 0.70 & 12
\\
MAX & 20 &15 & 0.0765 & 0.46 & 5
\\ \hline
\end{tabular}
\end{center}
\caption{Propagation parameters~\cite{Donato:2003xg}.}
\label{tab:paras}
\end{table}

The source term for DM annihilation/decay is given by
\begin{align}
Q(T,\vec{r}) = q(\vec{r}) \frac{dN_{\pbar}(T)}{dT}
\end{align}
where $dN_{\pbar}(T)/dT$ is the energy spectrum of the antiproton 
per one annihilation/decay, and $q(\vec{r})$ is given by
\begin{align}
q(\vec{r}) &= \frac{1}{2}\vev{\sigma v} \lrfp{\rho_{\text{DM}}(|\vec{r}|)}{m_{\text{DM}}}{2}
\qquad \text{for annihilating DM}\,,
\\
q(\vec{r}) &= \frac{1}{\tau_{\text{DM}}} \lrf{\rho_{\text{DM}}(|\vec{r}|)}{m_{\text{DM}}}
\qquad \text{for decaying DM}\,.
\end{align}
Here,  $m_{\text{DM}}$  and $\rho_{\text{DM}}(|\vec{r}|)$ are  the mass and the density profile of the DM, respectively. In addition, 
$\vev{\sigma v}$ is the annihilation cross section for the annihilating DM case, while
$\tau_{\text{DM}}$ is  the DM lifetime for the decaying DM case. 
 
\begin{figure}[t]
\begin{center}
\includegraphics[width=6.5cm]{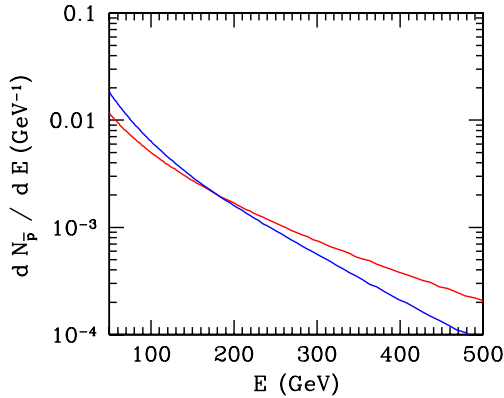}
\caption{Energy spectra of antiproton produced by
    non-relativistic annihilation of DM.  Red and blue lines
    correspond to those for $\chi\chi\rightarrow W^+W^-$ and
    $\bar{b}b$, respectively.  The DM mass is taken to be $2\
    {\rm TeV}$.
}
\label{fig:source}
\end{center}
\end{figure}

The differential equation (\ref{eq:diff}) can be solved analytically, which leads to
\begin{align}
\Phi_{\pbar}^{\text{DM}}(T) &= \frac{v(T)}{4\pi}\widetilde{G}(T) 
\frac{dN_{\pbar}(T)}{dT}\,.
\end{align}
For the source spectrum $dN_{\pbar}(T)/dT$, 
we consider the following DM annihilation and decay channels:
\begin{itemize}
\item annihilation: $\chi \chi \to W^+ W^-$,
\item decay:  $\chi  \to W^\pm \ell^\mp$,
\end{itemize}
where $\chi$ denotes the DM and $\ell$ is a charged lepton.
In Fig.~\ref{fig:source}, we show the numerical result for the
antiproton spectrum from $\chi \chi \to W^+ W^-$
obtained by Pythia 6.4~\cite{Sjostrand:2006za}. 
Our results agree well with the fitting formula in Ref.~\cite{Bergstrom:1999jc},
which we use in the following analysis.\footnote{
As for the fitting parameters $p_i(m_{\text{DM}})$ in Ref.~\cite{Bergstrom:1999jc}, 
we used $p_i(m_{\text{DM}}=5~\TeV)$ for $m_{\text{DM}}\ge 5~\TeV$.}
The spectrum from the decaying DM 
 with a mass of $m_{\text{DM}}$ 
is given by that from the annihilating DM 
with a mass of $m_{\text{DM}}/2$ 
rescaled by the factor of $1/2$.

In Fig.~\ref{fig:source}, we also show the antiproton spectrum from $\chi\chi \to b\bar{b}$.
The source spectra for $W^+W^-$ and $b\bar{b}$ are relatively close
in the parameter range of our interest.
We have checked that the resultant 
antiproton flux from $\chi\chi \to b\bar{b}$ is similar to the one from $\chi \chi \to W^+ W^-$.

The analytic expression for $\widetilde{G}(T)$ is given by~\cite{Maurin:2002ua}
\begin{align}
\widetilde{G}(T)&=\sum_{i=1}^\infty
\exp\lrf{-V_\text{c}L}{2K(T)}
\frac{y_i(T)}{A_i(T)\sinh(S_i(T)L/2)}
J_0\lrf{\zeta_i r_\odot}{R}
\end{align}
where
\begin{align}
y_i(T)&=\frac{4}{J_1^2(\zeta_i)R^2}
\int^R_0 r'dr' J_0\lrf{\zeta_i r'}{R}
\int^L_0 dz' \exp\lrf{V_\text{c}(L-z')}{2K(T)}
\sinh\lrf{S_i(L-z')}{2}
q(\vec{r})\,,
\nonumber\\
A_i(T)&=2h\Gamma_\text{ann}(T) + V_\text{c} 
+ K(T)S_i(T)\coth\lrf{S_i(T)L}{2}\,,
\nonumber\\
S_i(T)&=\sqrt{\frac{V_\text{c}^2}{K(T)^2} + \frac{4\zeta_i^2}{R^2}}\,.
\nonumber
\end{align}
Here, 
$J_0$ and $J_1$ are the zeroth and first order Bessel functions of the first kind, respectively, 
and $\zeta_i$ are the successive zeros of $J_0$.
We have calculated the Green's function $\widetilde{G}(T)$
by using the NFW density profile~\cite{Navarro:1996gj}
\begin{align}
\rho_{\text{DM}}(|\vec{r}|) &= \rho_{\odot}\frac{r_\odot}{|\vec{r}|}\left(\frac{1+r_\odot/r_s}{1+|\vec{r}|/r_s}\right)^2,
\end{align}
with $\rho_\odot=0.3$\,GeV/cm$^3$, $r_\odot=8.5$\,kpc and $r_s=20$\,kpc.
We parametrize the result as
\begin{align}
\widetilde{G}(T) &=\frac{1}{2}\vev{\sigma v}\frac{\rho_\odot^2}{m_\text{DM}^2} G_{\rm ann}(T)
\qquad \text{annihilating DM},
\\
\widetilde{G}(T) &=\frac{\rho_\odot}{m_\text{DM}\tau_{\text{DM}}} G_{\rm dec}(T)
\qquad \text{decaying DM}.
\end{align}
For annihilating DM, our numerical result of $G_{\rm ann}(T)$ 
agrees well with the fitting formula given in Ref.~\cite{Cirelli:2008id}.
For decaying DM, our numerical result is well reproduced with 
the following fitting function
\begin{align}
G_{\rm dec}(T) &= \exp\left(a_0+a_1\tau+a_2\tau^2+a_3\tau^3\right)
\times 10^{14}\text{sec},
\end{align}
where
\begin{align}
\tau=\log_{10}(T/\text{GeV}),
\end{align}
and the coefficients are shown in Table~\ref{tab:decay-fit} for MIN, MED, and MAX propagation models.\footnote{Our result slightly differs from the fitting formula presented in Ref.~\cite{Ibarra:2008qg} for high energy region, $T\gtrsim \mathcal O(100)$\,GeV.}

\begin{table}[t]
\begin{center}
\begin{tabular}{|c|cccc|}
\hline
Model & $a_0$ & $a_1$ & $a_2$ & $a_3$
\\
\hline
MIN & 1.1127 & 1.7495 & $-1.2730$ & 0.1412
\\
MED & 3.0662 & 0.8814 & $-0.8377$ & 0.09178
\\
MAX & 4.5815 & $-0.3546$ & $-0.2322$ & 0.02524
\\ \hline
\end{tabular}
\end{center}
\caption{fitting paramaters for decaying DM}
\label{tab:decay-fit}
\end{table}

\begin{figure}
\begin{center}
\includegraphics[height=14cm]{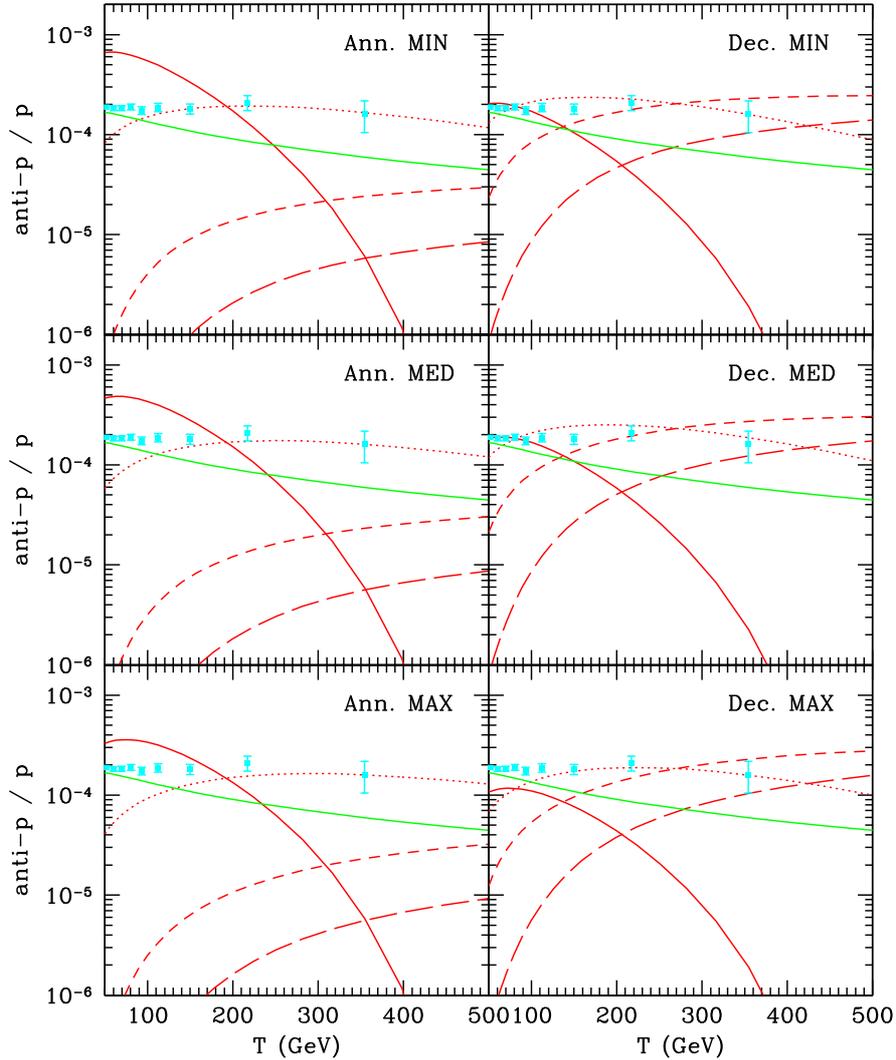}
\caption{The antiproton to proton ratio for MIN (top), MED (middle) 
and MAX (bottom) propagation models.
  The red lines
    are those predicted by the DM annihilation (left) and
    decay (right).  For the annihilation case, the DM mass is
    taken to be $0.5$ (solid), $2$ (dotted), $10$ (dashed), and $20\
    {\rm TeV}$ (long-dashed), while the annihilation cross section is
    taken to be $2\times 10^{-23}$, $2\times 10^{-24}$, and $6\times
    10^{-25}\ {\rm cm^3/sec}$, for MIN, MED, and MAX propagation
    models, respectively.  For the decay case, the DM mass is
    taken to be $1$ (solid), $3$ (dotted), $10$ (dashed), and $30\ {\rm TeV}$ (long-dashed), while the
    lifetime is $1\times 10^{26}$, $5\times 10^{26}$, and $2\times
    10^{27}\ {\rm sec}$, for MIN, MED, and MAX propagation models,
    respectively.
    The background is shown in the green line, and 
    the AMS-02 data are shown by the cyan points.}
\label{fig:generic}
\end{center}
\end{figure}

In Fig.~\ref{fig:generic}, we show our numerical results of $\pbar/p$ ratio for annihilating and decaying DM. 
Here, for the proton flux, we adopt the following fitting formula
\begin{align}
\frac{\Phi_{p}(T)}
{\text{m}^{-2}\text{sr}^{-1}\text{sec}^{-1}\text{GeV}^{-1}}
&=
\left[
10.0
-
\theta(-\tau_{300})
3.0\; \tau_{300}
-\theta(\tau_{300})
0.6\; \tau_{300}
\right]\times 10^3
\lrfp{T}{\text{GeV}}{-2.7},
\end{align}
where $\tau_{300}=\log_{10}(T/300~\text{GeV})$, 
which reproduces the newly released proton flux by AMS-02~\cite{AMS02} well 
for $T\gtrsim 30\GeV$.
In the figures, we also show the background spectrum represented 
by the fitting function in Ref.~\cite{Cirelli:2008id}:
\begin{align}
\log_{10}
\lrf{\Phi_{\pbar}^{\text{bkg}}}{\text{m}^{-2}\text{sr}^{-1}\text{sec}^{-1}\text{GeV}^{-1}}
&=
-1.64 + 0.07\tau -\tau^2-0.02\tau^3+0.028\tau^4.
\end{align}
As can be seen from the figures, the antiproton flux can be explained by annihilating/decaying DM with masses of ${\cal O}$(TeV).
Notice that the cross sections and lifetimes used in Fig.~\ref{fig:generic} are not the best-fit values, but just for presentation. The signal antiproton flux is proportional to $\vev{\sigma v}$ or $\tau_{\text{DM}}^{-1}$, and hence the AMS-02 antiproton flux can be explained in a wide range of DM mass if $\vev{\sigma v}$ or $\tau_{\text{DM}}$ is appropriately chosen.

\section{Implications for supersymmetric DM}

In this section, we briefly discuss the implications for some of SUSY DM candidates:
 annihilating Wino DM and decaying gravitino DM.

\subsection{Wino Dark Matter}

As shown in the previous section, if the antiproton flux is from the annihilating DM, it requires a mass of ${\cal O}$(TeV) and relatively large annihilation cross section of ${\cal O}(10^{-24}\,{\rm cm^3s^{-1}})$. 
Such a parameter space is natural in the Wino DM scenario.
In this case, the DM annihilation cross section is determined by its mass, and hence the antiproton flux depends only on the Wino mass.

\begin{figure}
\begin{center}
\includegraphics[height=14cm]{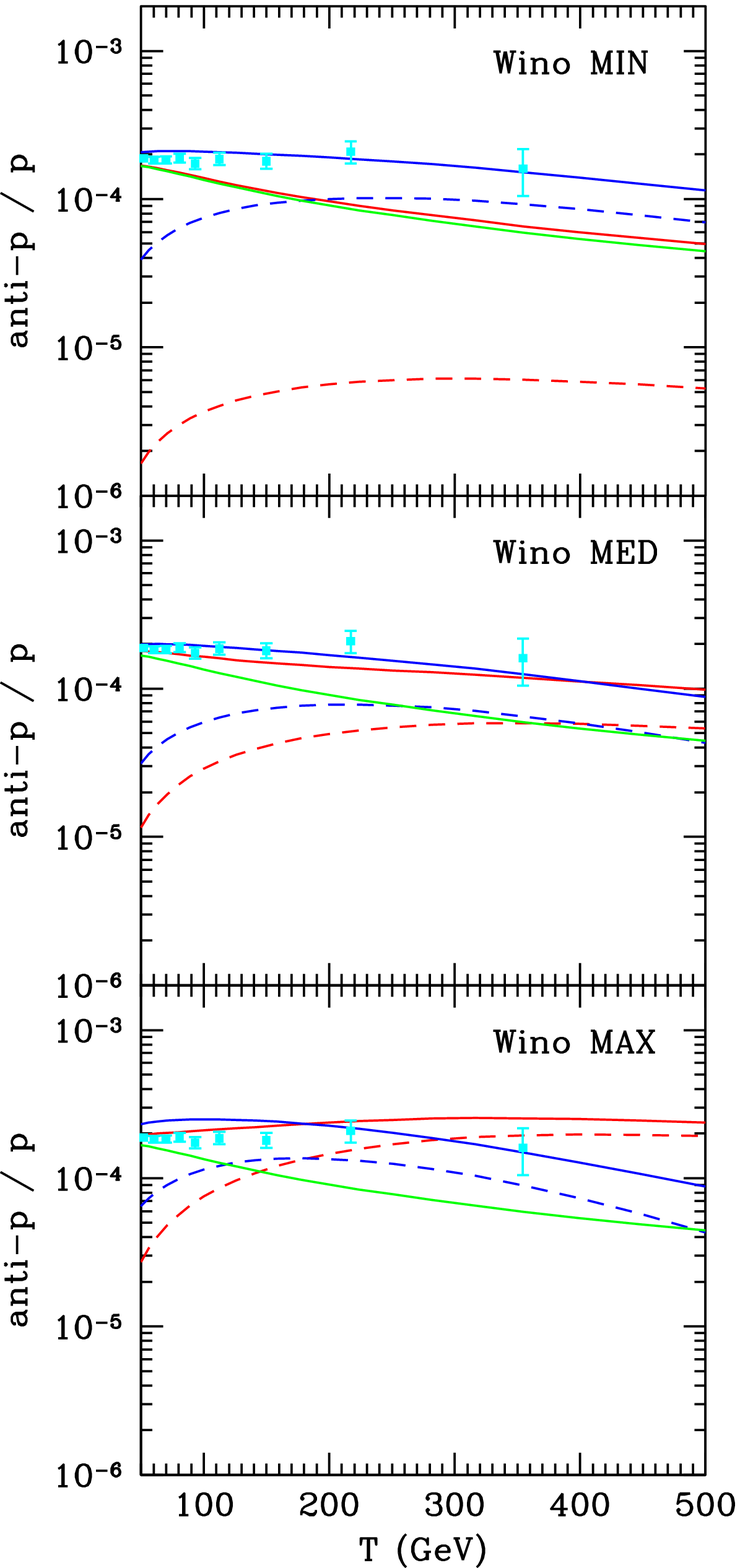}
\caption{The antiproton to proton ratio in  the Wino DM scenario
    for MIN (top), MED (middle) and MAX (bottom) propagation models.
Red lines are those for the Wino mass of $2.9\ {\rm TeV}$,
    while blue lines are those for the Wino mass of $2.2\ {\rm TeV}$
    (MIN), $1.7\ {\rm TeV}$ (MED), and $1.2\ {\rm TeV}$ (MAX).  The
    solid lines are signal plus background, while the dashed lines are
    signal-only.  The background is shown in the green line, and 
    the AMS-02 data are shown by the cyan points.
}
\label{fig:Wino}
\end{center}
\end{figure}

In Fig.~\ref{fig:Wino}, we show the $\pbar/p$ ratio for several Wino masses for MIN, MED and MAX propagation models. Here, we adopt the annihilation cross section in Ref.~\cite{Hisano:2004ds,Hisano:2008ti}.
As seen in the figure, Wino DM can account for the observed $\bar p/p$ ratio for several choices of masses.
Interestingly, for the MED and MAX propagation models, Wino mass of 2.9\,TeV can fit the antiproton data. Such a Wino mass is indeed predicted by the thermal relic abundance of Wino DM~\cite{Hisano:2006nn,Cirelli:2007xd}. As can be seen in the figures, smaller Wino masses can also explain the data.
In this case, a non-thermal production of Wino DM is necessary, such as gravitino~\cite{Moroi:1999zb,Gherghetta:1999sw,Ibe:2004gh,Ibe:2011aa}, moduli~\cite{Moroi:1999zb,Endo:2006ix}, or Q-ball decay~\cite{Fujii:2001xp}.
The antiproton flux at the AMS-02 may be the first hint of the Wino DM.

\subsection{Gravitino DM with R-parity violation}

In the scenario of gravitino DM with an R-parity violation~\cite{Takayama:2000uz,Buchmuller:2007ui}, a small R-parity violating coupling
and the Planck-suppressed interaction lead to a long DM lifetime. In the case of bilinear R-parity violation, its lifetime is given by~\cite{Takayama:2000uz,Buchmuller:2007ui}
\begin{align}
\tau_{3/2}\simeq 10^{26}\,\text{sec}
\lrfp{\lambda}{10^{-7}}{-2}
\lrfp{m_{3/2}}{1~\TeV}{-3},
\end{align}
where $m_{3/2}$ is the gravitino mass and $\lambda$ is the R-parity violating effective coupling.

As shown in Fig.~\ref{fig:generic}, if the high energy antiproton flux is from decaying gravitino DM, 
it implies $m_{3/2}\gtrsim \mathcal O(1)~\TeV$ and $\tau_{3/2}\simeq 10^{26}-10^{27}\,{\rm s}$.
This corresponds to an R-parity violating coupling of $\lambda\sim 10^{-7}-10^{-8}$.
Such a value of R-parity violation is attractive from the cosmological point of view, since it is large enough to solve the constraint from the long-lived next-to-lightest SUSY particle (NLSP) while small enough to avoid the baryon erasure in combination with sphaleron (cf.~\cite{Endo:2009cv}).

The gravitino abundance is given by~\cite{Moroi:1993mb,Bolz:2000fu,Pradler:2006qh}\footnote{Note that the NLSP quickly decays with R-parity violation without producing gravitinos.}
\begin{align}
\Omega_{3/2}h^2\simeq 0.1g_s^2\ln\left( \frac{1.3}{g_s} \right)\left(1+ \frac{m_{\tilde g}^2}{3m_{3/2}^2} \right)
\left( \frac{m_{3/2}}{1\,{\rm TeV}} \right)\left( \frac{T_{\rm R}}{10^9\,{\rm GeV}} \right),
\end{align}
where $g_s$ denotes the SU(3) gauge coupling constant, $m_{\tilde g}$ the gluino mass and $T_{\rm R}$ the reheating temperature after inflation.
Therefore, for explaining the AMS-02 antiprotons with gravitino DM of $m_{3/2} \gtrsim$ (a few) TeV, we need $T_{\rm R} \lesssim 10^9$\,GeV.
This is slightly lower than the temperature required by the standard thermal leptogenesis scenario~\cite{Fukugita:1986hr,Giudice:2003jh,Buchmuller:2004nz},
while non-thermal leptogenesis~\cite{Asaka:1999yd} can explain the observed baryon asymmetry.

\section{Discussion}

Now let us discuss various observational constraints on the annihilating/decaying DM scenario for explaining the AMS-02 antiprotons.
\begin{itemize}
\item Cosmic microwave background (CMB):
Recent Planck observation on the cosmic microwave background anisotropy constrains the DM annihilation cross section as
$f_{\rm eff} \langle\sigma v\rangle /m_{\rm DM} \lesssim 4\times 10^{-28}\,{\rm cm^3s^{-1}GeV^{-1}}$~\cite{Planck:2015xua},
where $f_{\rm eff}\sim 0.3$ (in the case of DM annihilation into $W^+W^-$ or $b\bar b$) is the effective fraction of the energy per DM annihilation that ionizes the hydrogen at the epoch of recombination~\cite{Padmanabhan:2005es}.
This leads to the following constraint
\begin{equation}
	\langle\sigma v\rangle \lesssim 1\times 10^{-24}\,{\rm cm^3s^{-1}} \left( \frac{m_{\rm DM}}{1\,{\rm TeV}} \right).
\end{equation}
For the MAX and MED model, this constraint is satisfied. For the MIN model, the required cross section is too large to satisfy this bound.
This does not constrain the decaying DM model, because the energy injection around the recombination epoch is sufficiently small.
Big-bang nucleosynthesis also constrains the annihilating DM model~\cite{Jedamzik:2004ip,Hisano:2008ti,Hisano:2009rc},
but the constraint is weaker than that from CMB.

\item Cosmic-ray positron: 
The annihilating/decaying DM also yields high-energy cosmic-ray positrons.
We have explicitly checked that the typical annihilating/decaying DM models explaining the AMS-02 antiprotons
do not conflict with the PAMELA~\cite{Adriani:2013uda} and AMS-02~\cite{Aguilar:2014} positron measurements for the MED and MAX models.

In the case of gravitino DM decaying into $W^\pm e^\mp$ or $W^\pm \mu^\mp$, 
a sizable positron flux is expected.
In particular,  when the gravitino decays into $W^\pm e^\mp$, a sharp edge of the positron spectrum at the energy around $m_{3/2}/2$ 
may be an interesting signature of this scenario.

\item Gamma-rays (continuum): 
Fermi satellite searches for gamma-rays from dwarf spheroidal galaxies and
puts severe constraint on the DM annihilation cross section and decay rate.
The constraint reads $\langle\sigma v\rangle \lesssim (0.2-10)\times 10^{-24}\,{\rm cm^3s^{-1}}$
for DM mass of $1-10$\,TeV in the case of annihilating DM~\cite{Ackermann:2015zua}.
For decaying DM, the diffuse gamma-ray background gives stringent constraint~\cite{Ando:2015qda}
and the typical constraint reads $\tau_{\rm DM} \gtrsim 10^{27}\,{\rm s}$ for DM mass of $1-10$\,TeV.
For both annihilating and decaying DM cases to explain the AMS-02 antiproton data, 
the MED and MAX models can satisfy the constraint.

\item Gamma-rays (line):
Annihilating or decaying DM
also produces line gamma. For example, the Wino DM annihilates into $Z\gamma$.
For DM mass of $\mathcal O$(TeV), the HESS telescope gives stringent constraint on such line signals
from the Galactic center in the case of DM annihilation~\cite{Abramowski:2013ax}.
However, the constraint significantly depends on the DM density profile.
According to Ref.~\cite{Baumgart:2014saa}, in which constraint from the gamma-ray line from Galactic center is derived for the case of Wino DM,
MED and MAX parameters are allowed for mild coring of the DM density profile around the Galactic center.

\end{itemize}

In summary, we studied the cosmic-ray antiproton flux from DM annihilation and decay
in light of the recent AMS-02 result.
It is possible to explain the observed $\bar p/p$ ratio by adding DM contributions to the typical astrophysical background.
In particular, we found that Wino DM with mass of around 3\,TeV can successfully account for the antiproton data,
which is consistent with the present DM abundance in the standard thermal freezeout scenario.
Decaying gravitino DM heavier than a few TeV can also explain the data.

\section*{Acknowledgement}
This work was supported by Grant-in-Aid for Scientific research No. 23104008 (TM), 26104001 (KH), 26104009 (KH and KN), 26247038 (KH), 
26247042 (KN), 26400239 (TM), 26800121 (KN), 26800123 (KH), and also by World Premier International Research Center Initiative (WPI Initiative), MEXT, Japan.

\section*{Note added}

While finalizing this manuscript, Ref.~\cite{Ibe:2015tma} appeared which has some overlaps with the present work.

\end{document}